\documentstyle[12pt,preprint,aps]{revtex}
\begin{document}
\preprint{CfA No. 4356}

\title{Variational calculations of dispersion coefficients
for interactions between H, He, and Li atoms}
\author{Zong-Chao Yan, James F. Babb, and A. Dalgarno}
\address{Harvard-Smithsonian Center for Astrophysics,
     60 Garden Street, Cambridge, Massachusetts 02138}
\author{G. W. F. Drake} 
\address {Department of Physics, University of Windsor, Windsor, 
Ontario, Canada N9B 3P4} 
\date{4 June, 1996} 
\maketitle

\begin{abstract}

The dispersion coefficients $C_6$,
$C_8$, and $C_{10}$ for the interactions between H, He, and Li
are calculated using variational wave functions
in Hylleraas basis sets with multiple exponential scale factors.
With these highly correlated wave functions,
significant improvements are made
upon previous calculations and our results
provide definitive values for these
coefficients.

\end{abstract}


\section {Introduction} 
At large separations $R$, the interaction potential
between two neutral atoms can be expressed in terms of
inverse powers of $R$, with the leading term $R^{-6}$
\cite{dalgarno1,dalgarno2}.
The nature of the long-range interaction can be described 
by the mutual perturbations of instantaneous multipoles 
of individual atoms. The coefficient of the 
$R^{-6}$ term comes from an instantaneous
dipole-dipole interaction
and the coefficient of the $R^{-8}$ term from an instantaneous
dipole-quadrupole interaction.

The precise evaluation of the dispersion coefficients
between atoms is computationally challenging, 
because it requires a summation over all
intermediate states, including the continuum. 
In actual calculations,
it is therefore essential to have an adequate representation
of the whole spectrum of the Hamiltonian. 
For atomic systems with more than one electron
the central problem is the inclusion of
electron-electron correlations.

Recently, significant progress \cite{drayan,yandra1} has
been made in variational 
calculations for the helium and lithium atoms 
using double and multiple basis sets in Hylleraas coordinates.
The nonrelativistic energies 
for helium have been obtained to better than
one part in $10^{16}$ for the entire singly-excited spectrum,
whereas the nonrelativistic energies for the low-lying
states of lithium are accurate 
to a few parts in $10^{11}$ to $10^{12}$
\cite{yandra2}. 
We have also performed a high precision calculation for
the lithium $2\,^2\!S\rightarrow 2\,^2\!P$
oscillator strength \cite{yanosi} 
which has been a subject of controversy 
for many years. Although there have been many calculations
for the dispersion coefficients between H, He, and Li
(see for example, 
\cite{dalgarno2,bishop,marinescu,jamieson,marinescu1,chenchung}),
the results involving Li vary over a considerable range.

Due to the recent progress of ultracold collisions in both
theory and experiment \cite{weiner}, precise forms of
long-range interaction potentials between various atoms
become very important.
The purpose of this paper is to present the results
of variational calculations in Hylleraas basis sets using
multiple nonlinear parameters.
The use of our highly correlated wave 
functions will improve
upon previous calculations and provide more definitive
values for the dispersion coefficients.

The theory of long-range forces is outlined in Sec.\ \ref{form}.
The structures of the basis sets for one, two, and three electron 
atomic systems are presented in Sec.\ \ref{cal}. Our final results
are tabulated and comparisons are made with the various previous
calculations. 
In the appendix, 
a derivation is given of the dispersion coefficients
for the Li($S$)-Li($P$) system.

\section {Formulation}
\label{form}
\subsection{Ground state dispersion coefficients}
In this section, we concentrate on interactions between 
atoms in their ground states. 
Using second-order perturbation theory,
the long-range part of 
interaction between two atoms a and b in their
ground states can be expanded
in terms of a series of inverse powers of the separation $R$
\cite{dalgarno1,dalgarno2}
\begin{eqnarray}
V_{\rm ab} &=&-\frac{C_6}{R^6}-\frac{C_8}{R^8}
-\frac{C_{10}}{R^{10}}\cdots\,,
\label{eq:a1}
\end{eqnarray}
where the coefficients $C_6$, $C_8$, and $C_{10}$ are 
\begin{eqnarray}
C_6 &=& \frac{3}{\pi}G_{\rm ab}(1,1)\,,
\label{eq:a2}
\end{eqnarray}
\begin{eqnarray}
C_8 &=& 
\frac{15}{2\pi}G_{\rm ab}(1,2)+\frac{15}{2\pi}G_{\rm ab}(2,1)\,,
\label{eq:a3}
\end{eqnarray}
\begin{eqnarray}
C_{10} &=& 
\frac{14}{\pi}G_{\rm ab}(1,3)+\frac{14}{\pi}G_{\rm ab}(3,1)
+\frac{35}{\pi}G_{\rm ab}(2,2)\,,
\label{eq:a4}
\end{eqnarray}
where
\begin{eqnarray}
G_{\rm ab}(l_{\rm a},l_{\rm b}) &=& \int_{0}^{\infty}
\alpha_{l_{\rm a}}^{\rm a}(i\omega)
\alpha_{l_{\rm b}}^{\rm b}(i\omega)d\omega\,.
\label{eq:a5}
\end{eqnarray}
In (\ref{eq:a5}), $\alpha_{l_{\rm a}}^{\rm a}(i\omega)$ is
the dynamic $2^{l_{\rm a}}$ polarizability for atom a at imaginary
frequency $i\omega$. 
The dynamic polarizability can be
expressed in terms of a sum 
over all intermediate states, including
the continuum (in atomic units throughout):
\begin{eqnarray}
\alpha_{l}(\omega) &=& \sum_{n}\frac{f_{n0}^{(l)}}{E_{n0}^2
-\omega^2}\,
\label{eq:a6}
\end{eqnarray}
with the $2^l$-pole oscillator strength
$f_{n0}^{(l)}$ being defined by
\begin{eqnarray}
f_{n0}^{(l)} &=& \frac{8\pi}{2l+1}E_{n0}|\langle\Psi_0
|\sum_{i} r_i^l Y_{lm}(\hat{\bf r}_i)|\Psi_n\rangle|^2\,,
\label{eq:a7}
\end{eqnarray}
where $E_{n0}=E_n-E_0$,
the sum $i$ runs over all the electrons in the atom,
$\Psi_0$ is the
ground state wave function, $E_0$ is the corresponding ground
state energy, and $\Psi_n$ is the $n$th intermediate
eigenfunction with the associated eigenvalue $E_n$.
An oscillator strength $\bar{f}_{n0}^{(l)}$ which is
independent of magnetic quantum number $m$ is obtained by
averaging over the initial state orientation degeneracy and
summing over the final state degeneracy. It is convenient
to introduce reduced matrix elements through the Wigner-Eckart
theorem \cite{edmonds}
\begin{eqnarray}
\langle \gamma' L'M'|\sum_{i} r_i^l Y_{lm}(\hat{\bf r}_i)|
\gamma L M\rangle &=& (-1)^{L'-M'}
{\left(\matrix{L'&l&L\cr{-M'}&m&M\cr}
\right)}\langle \gamma' L'||\sum_{i} r_i^l Y_{l}(\hat{\bf r}_i)||
\gamma L \rangle\,.
\label{eq:a8}
\end{eqnarray}
With the aid of a sum rule for the $3-j$ symbols, the 
oscillator strength can be written
\begin{eqnarray}
\bar{f}_{n0}^{(l)} &=& \frac{8\pi}{(2l+1)^2(2L_0+1)}
E_{n0}|\langle\Psi_0
||\sum_{i} r_i^l Y_{l}(\hat{\bf r}_i)||\Psi_n\rangle|^2\,,
\label{eq:a9}
\end{eqnarray}
where $L_0$ is the total angular momentum for the initial state.

Using the identity
\begin{eqnarray}
\frac{2}{\pi}\int_0^{\infty} 
\frac{d\omega}{(a^2+\omega^2)(b^2+\omega^2)}
&=&\frac{1}{ab(a+b)}\,, \  \ a,\ b>0 
\label{eq:a9a}
\end{eqnarray}
we can recast Eq.\ (\ref{eq:a5}) into the equivalent form 
\begin{eqnarray}
G_{\rm ab}(l_{\rm a},l_{\rm b}) &=&\frac{\pi}{2}\sum_{nn'}
\frac{f_{n0}^{(l_{\rm a})}f_{n'0}^{(l_{\rm b})}}
{ E_{n0}^{\rm a}  E_{n'0}^{\rm b}(E_{n0}^{\rm a}
+E_{n'0}^{\rm b})}\,,
\label{eq:a10}
\end{eqnarray}
where $E_{n0}^{\rm i}=E_n^{\rm i}-E_0^{\rm i}$ is
the excitation energy for atom i and is always positive 
for the atoms in the ground state. 
The procedure for evaluating $G_{\rm ab}(l_{\rm a},l_{\rm b})$ 
is to diagonalize Hamiltonian
in a basis set and sum over all intermediate
states directly according to (\ref{eq:a10}), and a
convergence study can be done by increasing the size of basis set
progressively.

The long-range part of the interaction between three ground state 
atoms is not exactly equal to the interaction energies
taken in pairs. There is an non-additive term which comes from
the third order perturbation. The leading terms in the expression
for the dispersion energy of the three-atom system are
\cite{dalgarno1,dalgarno2}
\begin{eqnarray}
V_{\rm abc} &=&-\frac{C_6^{\rm ab}}{r_{\rm ab}^6}
               -\frac{C_6^{\rm bc}}{r_{\rm bc}^6}
               -\frac{C_6^{\rm ca}}{r_{\rm ca}^6}
-\frac{\nu_{\rm abc}(3\cos\theta_{\rm a}\cos\theta_{\rm b}
\cos\theta_{\rm c}+1)}{(r_{\rm ab}r_{\rm bc}r_{\rm ca})^3}\,,
\label{eq:a12a}
\end{eqnarray}
where $\theta_{\rm a}$, $\theta_{\rm b}$, and $\theta_{\rm c}$
are the internal angles of the triangle formed by $r_{\rm ab}$, 
$r_{\rm bc}$ and $r_{\rm ca}$, and $\nu_{\rm abc}$ is the 
triple-dipole constant defined by
\begin{eqnarray}
\nu_{\rm abc} &=& \frac{3}{\pi}\int_0^{\infty}\alpha_1^{\rm a}
(i\omega)\alpha_1^{\rm b}(i\omega)\alpha_1^{\rm c}(i\omega)
d\omega\,.
\label{eq:a12}
\end{eqnarray}
\subsection{Excited state dispersion coefficients}
For two like atoms which are not both in their ground states,
the perturbation theory for calculating
the dispersion coefficients was given  by Marinescu and
Dalgarno \cite{marinescu1}. They worked out 
all the details for evaluating
the dispersion coefficients of alkali-metal dimers
in different excited states within a one-electron model potential
formalism. In this work, we examine the important case when one 
lithium atom is 
in the $2\,^2\!S$ ground state and the other lithium
atom is in 
the $2\,^2\!P$ state. A detailed derivation for many-electron
systems is given at the Appendix.

The zero-order wave function for the Li($S$)-Li($P$) system
can be written as a symmetrized product of two individual atomic
wave functions
\begin{eqnarray}
\Psi^{(0)} &=& \frac{1}{\sqrt 2}[\Psi_{\rm a}(L_1M_1;{\bf r})
\Psi_{\rm b}(L_2M_2;{\bf \rho})
+\beta \Psi_{\rm a}(L_1M_1;{\bf \rho})
\Psi_{\rm b}(L_2M_2;{\bf r})]\,,
\label {eq:a13}
\end{eqnarray}
where ${\bf r}$ and ${\bf \rho}$ 
represent all the internal coordinates
for the two atoms respectively,
$L_1$ and $L_2$ are their total orbital angular momenta,
$M_1$ and $M_2$ are the associated magnetic quantum numbers,
and $\beta=\pm 1$ describes
the symmetry due to Pauli exclusion principle. Following 
\cite{marinescu1},
first-order perturbation theory yields the interaction energy 
\begin{eqnarray}
V^{(1)}(L_2M_2;\beta) &=&  
-\frac{C_{2L_2+1}^{M_2\beta}}{R^{2L_2+1}}\,,
\label{eq:a14}
\end{eqnarray}
where
\begin{eqnarray}
C_{2L_2+1}^{M_2\beta} &=& 
\beta (-1)^{1+L_2+M_2}\frac{4\pi}{(2L_2+1)^2}
{{2L_2}\choose {L_2+M_2}}|\langle\Psi_{\rm a}
(0;{\bf r})||\sum_i r_i^{L_2}Y_{L_2}
(\hat{\bf r}_i)||\Psi_{\rm b}(L_2;{\bf r})\rangle|^2\,.
\label{eq:a15}
\end{eqnarray}
The first-order energy
correction is proportional to 
$\beta$. Therefore, for two unlike atoms
$(\beta=0)$ in the asymptotic
region where overlap between 
two atoms can be neglected, there is no
first-order correction to 
the interaction. For two like atoms, however,
there is a first-order 
correction to the interaction energy as long as
two atoms are in different angular momentum states. 
For the Li($S$)-Li($P$) system, 
the interaction is proportional to 
$R^{-3}$. The interaction 
between two ground state atoms is 
always attractive (see (\ref{eq:a1})) but
the interaction between two 
like atoms in different angular momentum states
is equally likely to be attractive and repulsive.

The leading energy correction obtained from the second-order 
perturbation theory for the Li($S$)-Li($P$) system is
\begin{eqnarray}
V^{(2)} &=& -\frac{C_6^{M_2}}{R^6}\,,
\label {eq:a16}
\end{eqnarray}
where
\begin{eqnarray}
C_6^{M_2} &=&\sum_{st}\frac{\Omega_{st}}
{E_{st}^{(0)}-E^{(0)}}\,
\label {eq:a17}
\end{eqnarray}
with
\begin{eqnarray}
\Omega_{st} &=&
|\langle
\Psi_{\rm a}(0;{\bf r})
||\sum_i r_i Y_1(\hat{\bf r}_i)||
\chi(1;{\bf r})
\rangle| ^2\nonumber\\
&\times&
\sum_{\lambda}
G(1,1,1,\lambda,1,M_2)
|\langle
\Psi_{\rm b}(1;{\bf \rho})
||\sum_j \rho_j Y_{1}(\hat{\bf \rho}_j)||
\omega(\lambda;{\bf \rho})
\rangle|^2\,.
\label {eq:a18}
\end{eqnarray}
In (\ref{eq:a17}), 
the summation should exclude one term which gives
rise to $E_{st}^{(0)}=E^{(0)}$. 
Note that $C_6^{M_2}$ is independent of $\beta$. 
The values of $G$ are listed in Table \ref{ag}.

It should be noted that 
the terms with $\lambda=1$ in (\ref{eq:a18}),
which corresponds to transitions 
between even parity states $P^{\rm e}$ 
and odd parity state $P^{\rm o}$,
are missing in one-electron model potential methods
\cite{marinescu1}. 
The dominant contribution comes from the radiative
transition between the lowest doubly excited doublet state
$1s2p2p\,^2P^{\rm e}$, which is stable against autoionization
\cite{buchet}, to the singly excited doublet state 
$1s^22p\,^2P^{\rm o}$. The contribution of the $\lambda=1$ term
to $C_6^{M_2}$ is $0.980\,82(5)$ for $M_2=\pm 1$, 
and $0.392\,32(2)$ for $M_2=0$.

\section {Calculations and results}
\label {cal}
For the hydrogen atom, the following Sturmian basis set 
\cite{hill} is used
to diagonalize the Hamiltonian:
\begin{eqnarray}
\{r^l e^{-\beta r/2}L_n^{(2l+2)}(\beta r)\}\,,
\label{eq:a20}
\end{eqnarray}
where $L_n^{(2l+2)}(\beta r)$ is the generalized
Laguerre polynomial and the parameter $\beta$ is chosen to be 
$\beta=2/(l+1)$. This basis set has proven to be numerically
stable as the size of basis set is enlarged. 

For the helium atom, the basis set is constructed using Hylleraas
coordinates \cite{drayan1}
\begin{eqnarray}
\{\chi_{ijk} &=& 
r_1^i\,r_2^j\,r_{12}^k\,e^{-\alpha r_1-\beta r_2}\}\,,
\label{eq:a21}
\end{eqnarray}
and the wave functions are expanded from doubled basis
sets.
The explicit form for the wave function is
\begin{eqnarray}
\Psi({\rm\bf r}_1,{\rm\bf r}_2)& =& \sum_{ijk}\,[a_{ijk}^{(1)}\,
\chi_{ijk}(\alpha_1,\beta_1)
+a_{ijk}^{(2)}\,\chi_{ijk}(\alpha_2,\beta_2)]
\pm\, {\rm exchange}\,,
\label{eq:a22}
\end{eqnarray}
and $i+j+k\le\Omega$. 
A complete optimization is then performed with
respect to the two sets of nonlinear parameters $\alpha_1$,
$\beta_1$, and 
$\alpha_2$, $\beta_2$. The screened hydrogenic wave
function is also included explicitly in the basis set. These
techniques yield much 
improved convergence relative to single basis
set calculations.

For the lithium atom, 
the basis set is also constructed in Hylleraas
coordinates \cite{yandra1}
\begin{eqnarray}
\{\phi_{t,\mu_t} (\alpha_t,\beta_t,\gamma_t) 
&=& r_1^{j_1}\,r_2^{j_2}
\,r_3^{j_3}\,r_{12}^{j_{12}}\,r_{23}^{j_{23}} 
\,r_{31}^{j_{31}}
\mbox{}e^{-\alpha_t r_1-\beta_t r_2-\gamma_t r_3}\}\,,
\label {eq:a23}
\end{eqnarray}
where
$\mu_t$ denotes a sextuple of integer powers $j_1$, $j_2$, $j_3$,
$j_{12}$, $j_{23}$, and $j_{31}$, 
index $t$ labels different sets of
nonlinear parameters $\alpha_t$, $\beta_t$ and $\gamma_t$.
Except for some truncations, all terms are included such that
\begin{eqnarray}
j_1+j_2+j_3+j_{12}+j_{23}+j_{31} &\leq& \Omega\,.
\label {eq:a24}
\end{eqnarray}
The wave function is expanded from the multiple basis sets
\begin{eqnarray}
\Psi ({\bf r}_1,{\bf r}_2,{\bf r}_3) &=& {\cal A} \sum_{t}
\sum_{\mu_t}a_{t,\mu_t}\phi_{t,\mu_t} (\alpha_t,\beta_t,\gamma_t)
\nonumber\\
&&\mbox{}\times ({\rm angular\ function})
({\rm spin\ function})\,.
\label {eq:aa25}
\end{eqnarray}
A similar optimization is also performed with respect to all the
nonlinear parameters. 

Table \ref{a} contains the values of the static polarizabilities
$\alpha_1(0)$, $\alpha_2(0)$, and $\alpha_3(0)$ for H, He, and Li
in their ground states. Tables \ref{e} and \ref{g} present
the comparison with selected previous calculations for He and
Li.  Using the Sturmian basis sets
containing up to 70 terms yields the well-known exact results
for the H atom. For He, the largest size of basis set for
the ground state is 504. For the intermediate states, the largest
sizes of basis sets are 728, 733, and 792 respectively for the
$P$, $D$, and $F$ symmetries. Table \ref{e} shows that
our value for $\alpha_1(0)$ is in perfect agreement 
with the best previous results of Bishop and Pipin \cite{bishop}
and Jamieson 
{\it et al.\/} \cite{jamieson} within the first 7 digits.
However, our value of $\alpha_1(0)$,
as well as $\alpha_2(0)$ and $\alpha_3(0)$,
has converged to several more significant 
figures, as indicated by the extrapolation uncertainty in
parentheses. For Li, with the fixed size of basis set 919
for the ground state,
Table \ref{d} contains the convergence
studies of $\alpha_1(0)$ in both length and velocity forms,
as the number of terms for the intermediate $P$ symmetry is
progressively increased. As a further numerical check,
we calculated $\alpha_1(0)$
for the Li atom by solving an inhomogeneous equation, using
the Dalgarno-Lewis method \cite{dalgarnolewis} as follows:
\begin{eqnarray}
\alpha_{l}(0) &=& -\frac{8\pi}{2l+1}\langle\Psi(2\,^2\!S)
|\sum_{i} r_i^l Y_{lm}(\hat{\bf r}_i)|\Psi_1\rangle\,,
\label{eq:a25}
\end{eqnarray}
where $\Psi_1$ satisfies the inhomogeneous equation
\begin{eqnarray}
(H_0-E_0)\Psi_1+(\sum_{i} r_i^l Y_{lm}(\hat{\bf r}_i)-E_1)
\Psi(2\,^2\!S) &=&0 \,.
\label{eq:a26}
\end{eqnarray}
In the above equation, $H_0$ is the unperturbed Li Hamiltonian,
$E_0$ is the Li ground state energy, and
\begin{eqnarray}
E_1 &=& \langle\Psi(2\,^2\!S)
|\sum_{i} r_i^l Y_{lm}(\hat{\bf r}_i)|\Psi(2\,^2\!S)\rangle\,.
\label{eq:a27}
\end{eqnarray}
Eq.\ (\ref{eq:a26}) is solved variationally by expanding
$\Psi_1$ in terms of a Hylleraas basis set. 
The two procedures are equivalent. Using basis sets
up to 2136, $\alpha_1(0)$ converges monotonically to $164.109\,8$.
Based on these convergence studies, our final extrapolated 
value of $\alpha_1(0)$ is $164.111(2)$. We have also done
similar convergence studies for $\alpha_2(0)$ and $\alpha_3(0)$.
Our results for $\alpha_1(0)$ and $\alpha_2(0)$ are in good 
accord with the results of Pipin and Bishop \cite{pipinb}.
The model potential results of Marinescu {\it et al.\/} 
\cite{marinescu} agree with the present calculations at
the $0.05\%$ and $0.1\%$ levels for $\alpha_2(0)$ and 
$\alpha_3(0)$ respectively.

Table \ref{b} shows our values of the two-body dispersion
coefficients $C_6$, $C_8$, and $C_{10}$ for the
ground state H, He, and Li atoms.
For the H-H case, these coefficients can be
calculated to arbitrarily high precision. Our value
of $C_6$ is in complete agreement with the value of
Margoliash and Meath \cite{margoliash}.
Comparisons with the previous calculations for 
He($1\,^1\!S$)-He($1\,^1\!S$) and Li($2\,^2\!S$)-Li($2\,^2\!S$)
are listed in Tables \ref{f} and \ref{h}.

For the He-He case,
our $C_6$ and $C_8$ are in excellent agreement with 
the results of Bishop and Pipin \cite{bishop}, but more
precise by about three
orders of magnitude. As for $C_{10}$, a small disagreement
of about 1 ppm exists. The value of Jamieson 
{\it et al.\/}\cite{jamieson} for $C_6$ also agrees with
our value. 

For the Li-Li case, convergence studies for $C_6$ in both length
and velocity forms are listed in Table \ref{d}. The agreement
between the two forms is satisfactory. From Table \ref{h},
it can be seen 
that the result of Stacey and Dalgarno \cite{stacey} 
for $C_6$ is in close agreement with the present calculation.
The model potential results of Marinescu {\it et al.\/}
\cite{marinescu} for $C_6$, $C_8$, and $C_{10}$ differ
from our calculations at the $0.1\%$ to $0.3\%$ level.

Table \ref{c} lists the triple-dipole constants 
$\nu_{\rm abc}$ for the combinations between
three ground state atoms H, He, and Li,
together with the previous values of Stacey and Dalgarno 
\cite{stacey}. The overall agreement is about $1\%$.

Finally, Tables \ref{d3} and \ref{i} list values of $C_3$
and $C_6$ and a comparison with the previous calculations
for the interaction between the ground state Li($2\,^2\!S$)
and the excited Li($2\,^2\!P$). $C_3$, which is proportional to
the square of the resonant dipole matrix element,
has recently been calculated to high precision \cite{yanosi}.
As for $C_6$, our values agree with the model potential
calculations of Marinescu and Dalgarno \cite{marinescu1}
at about the $0.3\%$ level.

\acknowledgments
Research support by the U.S. National Science Foundation, 
the Natural Sciences and Engineering Research
Council of Canada, and the U.S. Department of Energy,
Office of Basic Energy Sciences, is gratefully acknowledged.

\appendix
\section{}
In the appendix, we discuss the dispersion coefficients
for the Li($S$)-Li($P$) system. The zero-order wave function
for this system can be written as a symmetrized product of two
individual atomic wave functions
\begin{eqnarray}
\Psi^{(0)} &=& \frac{1}{\sqrt 2}[\Psi_{\rm a}(L_1M_1;{\bf r})
\Psi_{\rm b}(L_2M_2;{\bf \rho})
+\beta \Psi_{\rm a}(L_1M_1;{\bf \rho})
\Psi_{\rm b}(L_2M_2;{\bf r})]\,,
\label {eq:ap1}
\end{eqnarray}
where ${\bf r}$ and 
${\bf \rho}$ represent all the internal coordinates
for the two atoms respectively, $L_1$ and $L_2$ are
their total orbital angular momenta, 
$M_1$ and $M_2$ are the associated magnetic quantum numbers,
and $\beta=\pm 1$ describes
the symmetry due to Pauli exclusion principle. Following
\cite{marinescu1}, 
the interaction potential for two neutral atoms is
\begin{eqnarray}
V &=& \sum_{l=1}^{\infty}\sum_{L=1}^{\infty}
\frac{V_{lL}}{R^{l+L+1}}\,,
\label {eq:ap2}
\end{eqnarray}
where
\begin{eqnarray}
V_{lL} &=& (-1)^L 4\pi (l,L)^{-1/2}\sum_{ij}\sum_{\mu}
K_{lL}^{\mu}r_i^l \rho_j^L Y_{l\mu}(\hat{\bf r}_i)
Y_{L-\mu}(\hat{\bf \rho}_j)\,.
\label {eq:ap3}
\end{eqnarray}
In the above equation, $(l,L,\ldots)=(2l+1)(2L+1)\ldots$,
and
\begin{eqnarray}
K_{lL}^{\mu} &=& \left[ { {l+L}\choose {l+\mu} }
{ {l+L}\choose {L+\mu} }\right]^{1/2}\,.
\label {eq:ap4}
\end{eqnarray}

\subsection{The first-order energy}
The first-order energy is given by
\begin{eqnarray}
V^{(1)} &=& \frac{1}{2} A_1+\frac{1}{2} A_2+\beta A_3\,
\label {eq:app4}
\end{eqnarray}
with
\begin{eqnarray}
A_1  &=&
\langle
\Psi_{\rm a}(L_1M_1;{\bf r})
\Psi_{\rm b}(L_2M_2;{\bf \rho})|V|
\Psi_{\rm a}(L_1M_1;{\bf r})
\Psi_{\rm b}(L_2M_2;{\bf \rho})
\rangle \,,
\label {eq:ap5}
\end{eqnarray}
\begin{eqnarray}
A_2  &=&
\langle
\Psi_{\rm a}(L_1M_1;{\bf \rho})
\Psi_{\rm b}(L_2M_2;{\bf r})|V|
\Psi_{\rm a}(L_1M_1;{\bf \rho})
\Psi_{\rm b}(L_2M_2;{\bf r})
\rangle \,,
\label {eq:ap6}
\end{eqnarray}
\begin{eqnarray}
A_3  &=&
\langle
\Psi_{\rm a}(L_1M_1;{\bf r})
\Psi_{\rm b}(L_2M_2;{\bf \rho})|V|
\Psi_{\rm a}(L_1M_1;{\bf \rho})
\Psi_{\rm b}(L_2M_2;{\bf r})
\rangle \,.
\label {eq:ap7}
\end{eqnarray}
Substituting (\ref{eq:ap2}) into (\ref{eq:ap5}), one has
\begin{eqnarray}
A_1 &=& \sum_{lL\mu} \frac{(-1)^L 4\pi}{R^{l+L+1}}
(l,L)^{-{1/2}}K_{lL}^{\mu}
\langle
\Psi_{\rm a}(L_1M_1;{\bf r})
|\sum_i r_i^l Y_{l\mu}(\hat{\bf r}_i)|
\Psi_{\rm a}(L_1M_1;{\bf r})
\rangle\nonumber \\
&&\times
\langle
\Psi_{\rm b}(L_2M_2;{\bf \rho})
|\sum_j \rho_j^L Y_{L-\mu}(\hat{\bf \rho}_j)|
\Psi_{\rm b}(L_2M_2;{\bf \rho})
\rangle\,.
\label {eq:ap8}
\end{eqnarray}
>From the Wigner-Eckart theorem (\ref{eq:a8}), one has
\begin{eqnarray}
&&\langle
\Psi_{\rm a}(L_1M_1;{\bf r})
|\sum_i r_i^l Y_{l\mu}(\hat{\bf r}_i)|
\Psi_{\rm a}(L_1M_1;{\bf r})
\rangle\nonumber \\
&&=
(-1)^{L_1-M_1}
{\left(\matrix{L_1 & l & L_1\cr{-M_1}&\mu&M_1\cr} \right)}
\langle
\Psi_{\rm a}(L_1;{\bf r})
||\sum_i r_i^l Y_{l}(\hat{\bf r}_i)||
\Psi_{\rm a}(L_1;{\bf r})
\rangle\,.
\label {eq:ap9}
\end{eqnarray}
For $L_1=0$, the $3-j$ symbol is zero when $l\ge 1$.
Thus, $A_1=0$. Similarly, $A_2=0$.
For $A_3$, after using the Wigner-Eckart theorem, we obtain
\begin{eqnarray}
A_3 &=& \sum_{lL\mu} \frac{(-1)^L 4\pi}{R^{l+L+1}}
(l,L)^{-{1/2}}K_{lL}^{\mu}
\langle
\Psi_{\rm a}(L_1;{\bf r})
||\sum_i r_i^l Y_{l}(\hat{\bf r}_i)||
\Psi_{\rm b}(L_2;{\bf r})
\rangle\nonumber \\
&&\times
\langle
\Psi_{\rm b}(L_2;{\bf \rho})
||\sum_j \rho_j^L Y_{L}(\hat{\bf \rho}_j)||
\Psi_{\rm a}(L_1;{\bf \rho})
\rangle g\,,
\label {eq:ap10}
\end{eqnarray}
where, for $L_1=0$, 
\begin{eqnarray}
g &=& 
(-1)^{L_1-M_1}
{\left(\matrix{L_1&l&L_2\cr{-M_1}&\mu&M_2\cr}\right)}
(-1)^{L_2-M_2}
{\left(\matrix{L_2&L&L_1\cr{-M_2}&-\mu&M_1\cr}\right)}
\nonumber\\
&=&\frac{(-1)^{L_2-M_2}}{2L_2+1}\delta_{l,L_2}\delta_{L,L_2}
\delta_{\mu,-M_2}\,.
\label {eq:ap11}
\end{eqnarray}
Since \cite{edmonds}
\begin{eqnarray}
\langle L'||Y_l|| L\rangle &=& (-1)^{L'-L}
\langle L ||Y_l|| L'\rangle\,,
\label {eq:ap12}
\end{eqnarray}
we finally have
\begin{eqnarray}
V^{(1)}(L_2M_2;\beta) &=&  
-\frac{C_{2L_2+1}^{M_2\beta}}{R^{2L_2+1}}\,,
\label{eq:ap13}
\end{eqnarray}
where
\begin{eqnarray}
C_{2L_2+1}^{M_2\beta} &=& 
\beta (-1)^{1+L_2+M_2}\frac{4\pi}{(2L_2+1)^2}
{{2L_2}\choose {L_2+M_2}}|\langle\Psi_{\rm a}
(0;{\bf r})||\sum_i r_i^{L_2}Y_{L_2}
(\hat{\bf r}_i)||\Psi_{\rm b}(L_2;{\bf r})\rangle|^2\,.
\label{eq:ap14}
\end{eqnarray}

\subsection{The second-order energy}
Let the complete set of the system be
\begin{eqnarray}
\{\chi(L_sM_s;{\bf r})\omega(L_tM_t;{\bf \rho})\}
\label{eq:ap15}
\end{eqnarray}
with the energy eigenvalue $E_{st}^{(0)}=E_s^{(0)}+E_t^{(0)}$.
The energy for the unperturbed system is 
$E^{(0)}=E_1^{(0)}+E_2^{(0)}$.
According to the second-order perturbation theory,
the second-order energy is
\begin{eqnarray}
V^{(2)} &=& -\sum_{st} \frac{ |\langle\Psi^{(0)}|V|
\chi(L_sM_s;{\bf r})\omega(L_tM_t;{\bf \rho})\rangle|^2 }
{E_{st}^{(0)}-E^{(0)}}\nonumber\\
&=&-\sum_{st}\frac{T}{E_{st}^{(0)}-E^{(0)}}\,,
\label{eq:ap16}
\end{eqnarray}
where $T$ can be written as
\begin{eqnarray}
T &=& B_1+B_2+\beta B_3\,
\label{eq:ap17}
\end{eqnarray}
with
\begin{eqnarray}
B_1 &=&
\frac{1}{2}
\sum_{L_sM_s}\sum_{L_tM_t}
\langle
\Psi_{\rm a}(L_1M_1;{\bf r})
\Psi_{\rm b}(L_2M_2;{\bf \rho})|V|
\chi(L_sM_s;{\bf r})
\omega(L_tM_t;{\bf \rho})
\rangle^2 \,,
\label {eq:ap18}
\end{eqnarray}
\begin{eqnarray}
B_2 &=&
\frac{1}{2}
\sum_{L_sM_s}\sum_{L_tM_t}
\langle
\Psi_{\rm a}(L_1M_1;{\bf \rho})
\Psi_{\rm b}(L_2M_2;{\bf r})|V|
\chi(L_sM_s;{\bf r})
\omega(L_tM_t;{\bf \rho})
\rangle^2 \,,
\label {eq:ap19}
\end{eqnarray}
\begin{eqnarray}
B_3 &=&\sum_{L_sM_s}\sum_{L_tM_t}
\langle
\Psi_{\rm a}(L_1M_1;{\bf r})
\Psi_{\rm b}(L_2M_2;{\bf \rho})|V|
\chi(L_sM_s;{\bf r})
\omega(L_tM_t;{\bf \rho})
\rangle\nonumber\\
&\times&
\langle
\Psi_{\rm a}(L_1M_1;{\bf \rho})
\Psi_{\rm b}(L_2M_2;{\bf r})|V|
\chi(L_sM_s;{\bf r})
\omega(L_tM_t;{\bf \rho})
\rangle\,.
\label {eq:ap20}
\end{eqnarray}
After using the Wigner-Eckart theorem, we obtain
\begin{eqnarray}
B_1 &=& \frac{1}{2}\sum_{lL\mu}\sum_{l'L'\mu'}\sum_{L_sL_t}
\sum_{M_sM_t}
\frac{(4\pi)^2}{R^{l+L+l'+L'+2}}(-1)^{L+L'}
(l,L,l',L')^{-{1/2}}K_{lL}^{\mu}K_{l'L'}^{\mu'}\nonumber\\
&\times&
\langle
\Psi_{\rm a}(L_1;{\bf r})
||\sum_i r_i^l Y_{l}(\hat{\bf r}_i)||
\chi(L_s;{\bf r})
\rangle
\langle
\Psi_{\rm b}(L_2;{\bf \rho})
||\sum_j \rho_j^L Y_{L}(\hat{\bf \rho}_j)||
\omega(L_t;{\bf \rho})
\rangle\nonumber\\
&\times&
\langle
\Psi_{\rm a}(L_1;{\bf r})
||\sum_i r_i^{l'} Y_{l'}(\hat{\bf r}_i)||
\chi(L_s;{\bf r})
\rangle
\langle
\Psi_{\rm b}(L_2;{\bf \rho})
||\sum_j \rho_j^{L'} Y_{L'}(\hat{\bf \rho}_j)||
\omega(L_t;{\bf \rho})
\rangle\nonumber\\
&\times&
{\left(\matrix{L_1&l&L_s\cr{-M_1}&\mu&M_s\cr}\right)}
{\left(\matrix{L_2&L&L_t\cr{-M_2}&-\mu&M_t\cr}\right)}
{\left(\matrix{L_1&l'&L_s\cr{-M_1}&\mu'&M_s\cr}\right)}
{\left(\matrix{L_2&L'&L_t\cr{-M_2}&-\mu'&M_t\cr}\right)}\,.
\label {eq:ap21}
\end{eqnarray}
For $L_1=0$, the product of four $3-j$ symbols becomes
\begin{eqnarray}
\delta_{l,L_s}\delta_{l',L_s}
\delta_{\mu,-M_s}\delta_{\mu',-M_s}
\frac{1}{2L_s+1}
{\left(\matrix{L_2&L&L_t\cr{-M_2}&M_s&M_t\cr}\right)}
{\left(\matrix{L_2&L'&L_t\cr{-M_2}&M_s&M_t\cr}\right)}\,.
\label {eq:ap22}
\end{eqnarray}
Defining $G_1$ by
\begin{eqnarray}
G_1(L,L',L_s,L_t,L_2,M_2) &=&(-1)^{L+L'}\frac{(4\pi)^2}
{2(2L_s+1)^2}(L,L')^{-1/2}\sum_{M_sM_t}
K_{L_sL}^{-M_s}K_{L_sL'}^{-M_s}\nonumber\\
&\times&
{\left(\matrix{L_2&L&L_t\cr{-M_2}&M_s&M_t\cr}\right)}
{\left(\matrix{L_2&L'&L_t\cr{-M_2}&M_s&M_t\cr}\right)}\,,
\label {eq:ap23}
\end{eqnarray}
we have
\begin{eqnarray}
B_1 &=&\sum_{LL'L_sL_t}\frac{1}{R^{2L_s+L+L'+2}}
G_1(L,L',L_s,L_t,L_2,M_2)
|\langle
\Psi_{\rm a}(0;{\bf r})
||\sum_i r_i^{L_s} Y_{L_s}(\hat{\bf r}_i)||
\chi(L_s;{\bf r})
\rangle| ^2\nonumber\\
&\times&
\langle
\Psi_{\rm b}(L_2;{\bf \rho})
||\sum_j \rho_j^L Y_{L}(\hat{\bf \rho}_j)||
\omega(L_t;{\bf \rho})
\rangle
\langle
\Psi_{\rm b}(L_2;{\bf \rho})
||\sum_j \rho_j^{L'} Y_{L'}(\hat{\bf \rho}_j)||
\omega(L_t;{\bf \rho})
\rangle\,.
\label {eq:ap24}
\end{eqnarray}
Consider the leading term of $R^{-6}$. The only choice is
$L_s=1$, $L=1$, and $L'=1$. If another atom is in $L_2=1$
state, then $L_t=0$, 1, and 2. For this case,
\begin{eqnarray}
B_1 &=&\frac{1}{R^6}
|\langle
\Psi_{\rm a}(0;{\bf r})
||\sum_i r_i Y_1(\hat{\bf r}_i)||
\chi(1;{\bf r})
\rangle| ^2\nonumber\\
&\times&
\sum_{\lambda}
G_1(1,1,1,\lambda,1,M_2)
|\langle
\Psi_{\rm b}(1;{\bf \rho})
||\sum_j \rho_j Y_{1}(\hat{\bf \rho}_j)||
\omega(\lambda;{\bf \rho})
\rangle|^2\,.
\label {eq:ap25}
\end{eqnarray}
Similarly, for $B_2$ with $L_1=0$, we have
\begin{eqnarray}
B_2 &=&\sum_{ll'L_sL_t}\frac{1}{R^{2L_t+l+l'+2}}
G_2(l,l',L_t,L_s,L_2,M_2)
|\langle
\Psi_{\rm a}(0;{\bf \rho})
||\sum_j \rho_j^{L_t} Y_{L_t}(\hat{\bf \rho}_j)||
\omega(L_t;{\bf \rho})
\rangle| ^2\nonumber\\
&\times&
\langle
\Psi_{\rm b}(L_2;{\bf r})
||\sum_i r_i^l Y_{l}(\hat{\bf r}_i)||
\chi(L_s;{\bf r})
\rangle
\langle
\Psi_{\rm b}(L_2;{\bf r})
||\sum_i r_i^{l'} Y_{l'}(\hat{\bf r}_i)||
\chi(L_s;{\bf r})
\rangle\,
\label {eq:ap26}
\end{eqnarray}
with
\begin{eqnarray}
G_2(l,l',L_t,L_s,L_2,M_2) &=&\frac{(4\pi)^2}
{2(2L_t+1)^2}(l,l')^{-1/2}\sum_{M_sM_t}
K_{lL_t}^{M_t}K_{l'L_t}^{M_t}\nonumber\\
&\times&
{\left(\matrix{L_2&l&L_s\cr{-M_2}&M_t&M_s\cr}\right)}
{\left(\matrix{L_2&l'&L_s\cr{-M_2}&M_t&M_s\cr}\right)}\,.
\label {eq:ap27}
\end{eqnarray}
For $R^{-6}$, $L_t=1$, $l=1$, $l'=1$. Thus, 
for the case of $L_2=1$, $B_2$ becomes
\begin{eqnarray}
B_2 &=&\frac{1}{R^6}
|\langle
\Psi_{\rm a}(0;{\bf \rho})
||\sum_j \rho_j Y_1(\hat{\bf \rho}_j)||
\omega(1;{\bf \rho})
\rangle| ^2\nonumber\\
&\times&
\sum_{\lambda}
G_2(1,1,1,\lambda,1,M_2)
|\langle
\Psi_{\rm b}(1;{\bf r})
||\sum_i r_i Y_{1}(\hat{\bf r}_i)||
\chi(\lambda;{\bf r})
\rangle|^2\,.
\label {eq:ap28}
\end{eqnarray}

Finally, for $B_3$ with $L_1=0$, we have
\begin{eqnarray}
B_3 &=&\sum_{Ll'L_sL_t}\frac{1}{R^{L_s+L_t+L+l'+2}}
G_3(L,l',L_s,L_t,L_2,M_2) \nonumber\\
&\times&\langle
\Psi_{\rm a}(0;{\bf r})
||\sum_i r_i^{L_s} Y_{L_s}(\hat{\bf r}_i)||
\chi(L_s;{\bf r})
\rangle 
\langle
\Psi_{\rm a}(0;{\bf \rho})
||\sum_j \rho_j^{L_t} Y_{L_t}(\hat{\bf \rho}_j)||
\omega(L_t;{\bf \rho}) 
\rangle \nonumber\\
&\times&\langle
\Psi_{\rm b}(L_2;{\bf \rho})
||\sum_j \rho_j^{L} Y_{L}(\hat{\bf \rho}_j)||
\omega(L_t;{\bf \rho})
\rangle 
\langle
\Psi_{\rm b}(L_2;{\bf r})
||\sum_i r_i^{l'} Y_{l'}(\hat{\bf r}_i)||
\chi(L_s;{\bf r}) 
\rangle \,,
\label {eq:ap29}
\end{eqnarray}
with
\begin{eqnarray}
G_3(L,l',L_s,L_t,L_2,M_2) &=&(-1)^{L+Ls}\frac{(4\pi)^2}
{(2L_s+1)(2L_t+1)}(L,l')^{-1/2}\sum_{M_sM_t}
(-1)^{M_s+M_t}K_{L_sL}^{-M_s}K_{l'L_t}^{M_t}\nonumber\\
&\times&
{\left(\matrix{L_2&L&L_t\cr{-M_2}&M_s&M_t\cr}\right)}
{\left(\matrix{L_2&l'&L_s\cr{-M_2}&M_t&M_s\cr}\right)}\,.
\label {eq:ap30}
\end{eqnarray}
The only term which contributes $R^{-6}$ is the one with
$L_s=1$, $L_t=1$, $l'=1$, and $L=1$. For the case of $L_2=1$, 
one obtains
\begin{eqnarray}
B_3 &=& \frac{1}{R^6} G_3(1,1,1,1,1,M_2)\nonumber\\
&\times&\langle
\Psi_{\rm a}(0;{\bf r})
||\sum_i r_i Y_{1}(\hat{\bf r}_i)||
\chi(1;{\bf r})
\rangle
\langle
\Psi_{\rm a}(0;{\bf \rho})
||\sum_j \rho_j Y_{1}(\hat{\bf \rho}_j)||
\omega(1;{\bf \rho})
\rangle \nonumber\\
&\times&\langle
\Psi_{\rm b}(1;{\bf \rho})
||\sum_j \rho_j Y_{1}(\hat{\bf \rho}_j)||
\omega(1;{\bf \rho})
\rangle
\langle
\Psi_{\rm b}(1;{\bf r})
||\sum_i r_i Y_{1}(\hat{\bf r}_i)||
\chi(1;{\bf r})
\rangle \,.
\label {eq:ap31}
\end{eqnarray}
For the $S$ state $\Psi_{\rm a}(0;{\bf r})$, 
the parity is +1, and
for the $P$ state $\Psi_{\rm b}(1;{\bf r})$, 
the parity is $-1$.
Since these two states cannot be connected 
simultaneously to a third
parity eigenstate by a dipole operator, 
$B_3$ is therefore zero.

For two like atoms the spectra 
$\{\chi(LM;{\bf r})\}$ and $\{\omega(LM;{\bf \rho})\}$ 
are identical and $B_1$ and $B_2$ can be combined.
The final expression for the second-order energy correction to
the Li($S$)-Li($P$) system is
\begin{eqnarray}
V^{(2)} &=& -\frac{C_6^{M_2}}{R^6}\,,
\label {eq:ap32}
\end{eqnarray}
where 
\begin{eqnarray}
C_6^{M_2} &=&\sum_{st}\frac{\Omega_{st}}
{E_{st}^{(0)}-E^{(0)}}\,
\label {eq:ap33}
\end{eqnarray}
with
\begin{eqnarray}
\Omega_{st} &=&
|\langle
\Psi_{\rm a}(0;{\bf r})
||\sum_i r_i Y_1(\hat{\bf r}_i)||
\chi(1;{\bf r})
\rangle| ^2\nonumber\\
&\times&
\sum_{\lambda}
G(1,1,1,\lambda,1,M_2)
|\langle
\Psi_{\rm b}(1;{\bf \rho})
||\sum_j \rho_j Y_{1}(\hat{\bf \rho}_j)||
\omega(\lambda;{\bf \rho})
\rangle|^2\,.
\label {eq:ap34}
\end{eqnarray}
In (\ref{eq:ap34}), $G$ is defined by
\begin{eqnarray}
G(1,1,1,\lambda,1,M_2)&=&
G_1(1,1,1,\lambda,1,M_2)+G_2(1,1,1,\lambda,1,M_2)\,.
\label {eq:ap35}
\end{eqnarray}
It is easy to see that
\begin{eqnarray}
G_1(1,1,1,\lambda,1,M_2)&=&G_2(1,1,1,\lambda,1,M_2)\,.
\label {eq:ap36}
\end{eqnarray}
The algebraic coefficients $G(1,1,1,\lambda,1,M_2)$
are listed in Table \ref{ag}.

\begin{table}
\caption{The algebraic coefficients $G(1,1,1,\lambda,1,M_2)$.}
\label{ag}
\begin{tabular}{l c c c}
\multicolumn{1}{l}{}&
\multicolumn{1}{c}{$\lambda=0$}&
\multicolumn{1}{c}{$\lambda=1$}&
\multicolumn{1}{c}{$\lambda=2$}\\
\tableline
 $M_2=0$    & $\frac{64}{81}\pi^2$ 
 & $\frac{16}{81}\pi^2$ &$\frac{176}{405}\pi^2$\\
 $M_2=\pm 1$    & $\frac{16}{81}\pi^2$ 
 & $\frac{40}{81}\pi^2$ &$\frac{152}{405}\pi^2$
\end{tabular}
\end{table}

\begin{table}
\caption{Values of the static polarizabilities $\alpha_1(0)$,
$\alpha_2(0)$, and $\alpha_3(0)$ for
the ground state H, He, and Li atoms.}
\label{a}
\begin{tabular}{l c c c}
\multicolumn{1}{l}{System}&
\multicolumn{1}{c}{$\alpha_1(0)$}&
\multicolumn{1}{c}{$\alpha_2(0)$}&
\multicolumn{1}{c}{$\alpha_3(0)$}\\
\tableline
H    & 4.5 &15 &131.25\\
He   &1.383\,192\,174\,40(5) &2.445\,083\,101(2)
&10.620\,328\,6(2)\\
Li   &164.111(2) &1\,423.266(5) &39\,650.49(8)
\end{tabular}
\end{table}

\begin{table}
\caption{Comparison of static polarizabilities
$\alpha_1(0)$, $\alpha_2(0)$, and $\alpha_3(0)$
for He($1\,^1\!S$).}
\label{e}
\begin{tabular}{l c r@{}l r@{}l r@{}l}
\multicolumn{1}{l}{Author (year)}&
\multicolumn{1}{c}{Reference}&
\multicolumn{2}{c}{$\alpha_1(0)$}&
\multicolumn{2}{c}{$\alpha_2(0)$}&
\multicolumn{2}{c}{$\alpha_3(0)$}\\
\tableline
Luyckx {\it et al.\/} (77)& \cite{luyckx}&
1&.379& 2&.430 &10&.48\\
Thakkar (81)              & \cite{thakkar}&
1&.383\,12 &2&.443\,44 &10&.614\,4\\
Bishop and Pipin (93)     & \cite{bishop}&
1&.383\,192 &2&.445\,083 &10&.620\,360\\
Caffarel {\it et al.\/} (93) &\cite{caffarel}&
1&.382\,7 &2&.406\,6 &10&.36\\
Jamieson {\it et al.\/} (95)  & \cite{jamieson}&
1&.383\,192 && &&\\
Chen (95)                 & \cite{chen} &
1&.383\,32  && &&    \\
Chen and Chung (96)       & \cite{chenchung}&
1&.383\,27 &2&.445\,66 &10&.625\,2\\
This work   & &
1&.383\,192\,174\,40(5) &2&.445\,083\,101(2)
&10&.620\,328\,6(2)
\end{tabular}
\end{table}

\begin{table}
\caption{Comparison of static polarizabilities
$\alpha_1(0)$, $\alpha_2(0)$, and $\alpha_3(0)$
for Li($2\,^2\!S$).}
\label{g}
\begin{tabular}{l c r@{}l r@{}l r@{}l}
\multicolumn{1}{l}{Author (year)}&
\multicolumn{1}{c}{Reference}&
\multicolumn{2}{c}{$\alpha_1(0)$}&
\multicolumn{2}{c}{$\alpha_2(0)$}&
\multicolumn{2}{c}{$\alpha_3(0)$}\\
\tableline
Maeder and Kutzelnigg (79)        & \cite{maeder}& 
164&.3 &1\,383& &36\,795& \\
Muszy\'{n}ska {\it et al.\/} (82) & \cite{muszy}&
163&.8 && && \\
Pipin and Wo\'{z}nicki (83)       & \cite{pipinwo}&
163&.9 && &&\\
Pouchan and Bishop (84)           & \cite{pouchan}&
164&(2)  && &&\\
M\"{u}ller {\it et al.\/} (84)    & \cite{muller}&
163&.7 && &&\\
Knowles and Meath     (86)        & \cite{knowles}&
165&.8&  1\,486& & 36\,495& \\
Maroulis and Thakkar (89)         & \cite{maroulis}&
164&.5 &1\,428& && \\
Pipin and Bishop     (92)         & \cite{pipinb}&
164&.1 &1\,423& && \\
Ponomarenko and Shestakov (93)    & \cite{Pono}&
165&.2 && && \\
Marinescu {\it et al.\/} (94)     & \cite{marinescu}&
& &1\,424& &39\,688& \\
Wang and Chung       (94)         & \cite{wang}&
164&.08& & && \\
M\'{e}rawa {\it et al.\/} (94)    & \cite{merawa}&
164&.8& 1\,430& && \\
Kassimi and Thakkar       (94)    & \cite{kassimi}&
164&.2(1)& & && \\
Laughlin                  (95)    & \cite{laughlin}&
163&.91& &&& \\
This work                         &               &
164&.111(2) &1\,423&.266(5) &39\,650&.49(8)\\
Experiment (74)                   & \cite{molof}&
164&.0(3.4)& & && 
\end{tabular}
\end{table}

\begin{table}
\caption{Convergence of Li($2\,^2\!S$)
$\alpha_1(0)$ and Li($2\,^2\!S$)-Li($2\,^2\!S$) $C_6$
in length and velocity forms.}
\label{d}
\begin{tabular}{l r@{}l r@{}l r@{}l r@{}l}
\multicolumn{1}{c}{No. of terms}&
\multicolumn{2}{c}{$\alpha_1(0)$ (length)}&
\multicolumn{2}{c}{$\alpha_1(0)$ (velocity)}&
\multicolumn{2}{c}{$C_6$ (length)}&
\multicolumn{2}{c}{$C_6$ (velocity)}\\
\tableline
56  &164&.002 &165&.218 &1\,389&.76 &1\,409&.91\\
139 &164&.048 &164&.201 &1\,391&.21 &1\,393&.56\\
307 &164&.082 &164&.131 &1\,392&.37 &1\,393&.08\\
623 &164&.095 &164&.107 &1\,392&.80 &1\,392&.92\\
1175&164&.105 &164&.108 &1\,393&.17 &1\,393&.17\\
1846&164&.107 &164&.108 &1\,393&.23 &1\,393&.21 
\end{tabular}
\end{table}

\begin{table}
\caption{Values of $C_6$, $C_8$, and $C_{10}$ for two 
ground state atoms.}
\label{b}
\begin{tabular}{l c c c}
\multicolumn{1}{l}{System}&
\multicolumn{1}{c}{$C_6$}&
\multicolumn{1}{c}{$C_8$}&
\multicolumn{1}{c}{$C_{10}$}\\
\tableline
H-H &6.499\,026\,705\,405\,839\,313\,13 
&124.399\,083\,583\,622\,343\,609\,59
&3\,285.828\,414\,967\,421\,697\,872\,5\\   
He-He &1.460\,977\,837\,68(5) &14.117\,857\,340(5)
&183.691\,070\,5(7) \\
Li-Li &1\,393.39(16) &83\,425.8(4.2)& $73\,721(1)\times 10^2$\\
H-He &2.821\,343\,915\,28(6) &41.836\,376\,162(8)
&871.540\,471(1)\\
He-Li &22.507(1) &1\,083.16(5) &72\,602.1(1)\\
Li-H &66.536(5) &3\,279.99(2) &223\,016.6(5)  
\end{tabular}
\end{table}

\begin{table}
\caption{Comparison of $C_6$, $C_8$, and $C_{10}$
for the He($1\,^1\!S$)-He($1\,^1\!S$) system.}
\label{f}
\begin{tabular}{l c r@{}l r@{}l r@{}l}
\multicolumn{1}{l}{Author (year)}&
\multicolumn{1}{c}{Reference}&
\multicolumn{2}{c}{$C_6$}&
\multicolumn{2}{c}{$C_8$}&
\multicolumn{2}{c}{$C_{10}$}\\
\tableline
Luyckx {\it et al.\/} (77)& \cite{luyckx}&
1&.458   &14&.06   &182&.16\\
Glover and Weinhold (77)  & \cite{glover}&
1&.459\,7(55) && &&\\
Margoliash and Meath (78)& \cite{margoliash}&
1&.458    && &&\\
Bartolotti (80)           & \cite{bartolotti}&
1&.463\,8  & 14&.131  &183&.47\\
Thakkar (81)              & \cite{thakkar}&
1&.460\,82 &14&.111\,8  &183&.600\\ 
R\'{e}rat {\it et al.\/} (93) & \cite{rerat}&
1&.459\,3 &13&.883 && \\
Bishop and Pipin (93)     & \cite{bishop}&
1&.460\,977\,8 &14&.117\,855 &183&.691\,25\\
Jamieson {\it et al.\/} (95)  & \cite{jamieson}&
1&.460\,978 && &&\\
Chen (95)                 & \cite{mkchen} &
1&.461\,1   &14&.120   &183&.74\\
Chen and Chung (96)       & \cite{chenchung}&
1&.461\,06  &14&.120\,8  &183&.765\\
This work                &                 &
1&.460\,977\,837\,68(5) &14&.117\,857\,340(5)
&183&.691\,070\,5(7)
\end{tabular}
\end{table}

\begin{table}
\caption{Comparison of $C_6$, $C_8$, and $C_{10}$
for the Li($2\,^2\!S$)-Li($2\,^2\!S$) system.}
\label{h}
\begin{tabular}{l c r@{}l r@{}l r@{}l}
\multicolumn{1}{l}{Author (year)}&
\multicolumn{1}{c}{Reference}&
\multicolumn{2}{c}{$10^{-3}C_6$}&
\multicolumn{2}{c}{$10^{-4}C_8$}&
\multicolumn{2}{c}{$10^{-6}C_{10}$}\\
\tableline
Stacey and Dalgarno (68)          & \cite{stacey}&
1&.391  & & && \\
Manakov and Ovsiannikov (77)      & \cite{manakov}&
1&.360&& &&\\
Margoliash and Meath (78)         & \cite{margoliash}&
1&.387  & & && \\
Maeder and Kutzelnigg (79)        & \cite{maeder}&
1&.389   &8&.089  &6&.901\\
M\"{u}ller {\it et al.\/} (84)    & \cite{muller}&
1&.386  & & && \\
Bussery and Aubert-Fr\'{e}con (85)& \cite{bussery}&
1&.383 &7&.578\,3 &4&.816\,675 \\
Marinescu {\it et al.\/} (94)     & \cite{marinescu}&
1&.388 &8&.324 &7&.365\\
M\'{e}rawa {\it et al.\/} (94)    & \cite{merawa}&
1&.407\,8 &8&.431\,65 && \\
This work                         &              &
1&.393\,39(16) &8&.342\,58(42)& 7&.372\,1(1)
\end{tabular}
\end{table}

\begin{table}
\caption{Values of the triple-dipole constants 
$\nu_{\rm abc}$ for the three ground state atoms
H, He, and Li.}
\label{c}
\begin{tabular}{l c c}
\multicolumn{1}{l}{System}&
\multicolumn{1}{c}{$C_6$ (This work)}&
\multicolumn{1}{c}{$C_6$ (Ref.\ \cite{stacey})}\\
\tableline
H-H-H    &      21.642\,464\,510\,635\,978\,338\,11&\\
He-H-H   &       8.102\,240\,874\,3(2)&\\
He-He-H  &       3.268\,064\,896\,1(1)&\\
He-He-He &       1.479\,558\,606\,3(1)&\\
Li-H-H   &     275.979(7) & $276$\\
Li-He-H  &      89.830(5) & $89.6$\\
Li-He-He &      29.824(5) & $29.6$\\
Li-Li-H  &  6\,133.5(5)   & $6.12\times 10^3$\\
Li-Li-He &  1\,917.27(5)  & $1.91\times 10^3$\\
Li-Li-Li &  170\,595(6)     & $1.69\times 10^5$
\end{tabular}
\end{table}

\begin{table}
\caption{Values of $C_3$ and $C_6$ for the interaction 
between Li($2\,^2\!S$) and Li($2\,^2\!P$).} 
\label{d3}
\begin{tabular}{l r@{}l r@{}l r@{}l}
\multicolumn{1}{l}{$M_2$}&
\multicolumn{2}{c}{$\beta$}&
\multicolumn{2}{c}{$C_3$}&
\multicolumn{2}{c}{$C_6$}\\
\tableline
0 & 1& & 11&.000\,226(15) &2\,075&.05(5)\\
0 & --1& &--11&.000\,226(15)&2\,075&.05(5)\\
$\pm 1$ & 1& &--5&.500\,113\,3(74)&1\,406&.08(5)\\
$\pm 1$ & --1&&5&.500\,113\,3(74)&1\,406&.08(5)
\end{tabular}
\end{table}

\begin{table}
\caption{Comparison of $C_6$
for the Li($2\,^2\!S$)-Li($2\,^2\!P$) system.}
\label{i}
\begin{tabular}{l c r@{}l r@{}l}
\multicolumn{1}{l}{Author (year)}&
\multicolumn{1}{c}{Reference}&
\multicolumn{2}{c}{$C_6(M_2=0)$}&
\multicolumn{2}{c}{$C_6(M_2=\pm 1)$}\\
\tableline
Konowalov and Fish (83)           & \cite{konowalov}&
2\,100&(50) & 1\,750&(100) \\
Vign\'{e}-Maeder (84)             & \cite{vigne}&                 
2\,025& & 1\,374&  \\
Bussery and Aubert-Fr\'{e}con (85)& \cite{bussery}&
1\,927& & 1\,301&  \\
Marinescu and Dalgarno (95)       & \cite{marinescu1}&
2\,066& &1\,401& \\
This work                         &              &
2\,075&.05(5)& 1\,406&.08(5)
\end{tabular}
\end{table}


\begin{thebibliography}{99}
\bibitem{dalgarno1} A. Dalgarno and W. D. Davison, Adv. 
At.\ Mol.\ Phys.\  {\bf 2}, 1 (1966).
\bibitem{dalgarno2} A. Dalgarno, Adv.\ Chem.\ Phys.\ {\bf 12},
143 (1967).
\bibitem{drayan} G. W. F. Drake and Z.-C. Yan,
Phys.\ Rev.\ A {\bf 46}, 2378 (1992).
\bibitem{yandra1} Z.-C. Yan and G. W. F. Drake,
Phys.\ Rev.\ A {\bf 52}, 3711 (1995).
\bibitem{yandra2} Z.-C. Yan and G. W. F. Drake,
(unpublished).
\bibitem{yanosi} Z.-C. Yan and G. W. F. Drake,
Phys.\ Rev.\ A {\bf 52}, R4316 (1995).
\bibitem{bishop} D. M. Bishop and J. Pipin, 
Int.\ J.\ Quantum Chem.\ {\bf 45}, 349 (1993).
\bibitem{marinescu} M. Marinescu, H. R. Sadeghpour,
and A. Dalgarno, Phys.\ Rev.\ A {\bf 49}, 982 (1994).
\bibitem{jamieson} M. J. Jamieson, G. W. F. Drake,
and A. Dalgarno, Phys.\ Rev.\ A {\bf 51}, 3358 (1995).
\bibitem{marinescu1} M. Marinescu and A. Dalgarno, 
Phys.\ Rev.\ A {\bf 52}, 311 (1995).
\bibitem{chenchung} M.-K. Chen and K. T. Chung, 
Phys.\ Rev.\ A {\bf 53}, 1439 (1996).
\bibitem{weiner} J. Weiner, Adv. 
At.\ Mol.\ Opt.\ Phys.\  {\bf 35}, 45 (1995).
\bibitem{edmonds} A. R. Edmonds, {\it Angular Momentum in
Quantum Mechanics} 
(Princeton University Press, Princeton, 1985).
\bibitem{buchet} J. P. Buchet, M. C. Buchet-Poulizac,
H. G. Berry, and G. W. F. Drake,  Phys.\ Rev.\ A {\bf 7},
922 (1973).
\bibitem{hill} R. N. Hill, in {\it Atomic, Molecular, 
and Optical Physics Handbook}, edited by G. W. F. Drake
(American Institute of Physics, New York, 1996).
\bibitem{drayan1} G. W. F. Drake and Z.-C. Yan,
Chem.\ Phys.\ Lett.\ {\bf 229}, 486 (1994).
\bibitem{luyckx} R. Luyckx, Ph. Coulon, and H. N. W.
Lekkerkerker,
Chem.\ Phys.\ Lett.\ {\bf 48}, 187 (1977).
\bibitem{thakkar} A. J. Thakkar, J.\ Chem.\ Phys.\
{\bf 75}, 4496 (1981).
\bibitem{caffarel} M. Caffarel, M. R\'{e}rat, and
C. Pouchan, Phys.\  Rev.\ A {\bf 47}, 3704 (1993).
\bibitem{chen} M.-K. Chen, J.\ Phys.\ B {\bf 28}, 1349 (1995).
\bibitem{dalgarnolewis} A. Dalgarno and J. T. Lewis,
Proc. R. Soc. London Ser. A {\bf 233}, 70 (1955).
\bibitem{pipinb} J. Pipin and D. M. Bishop, 
Phys.\ Rev.\ A {\bf 45},
2736 (1992).
\bibitem{maeder} F. Maeder and W. Kutzelnigg, 
Chem.\ Phys.\ {\bf 42},
95 (1979).
\bibitem{muszy} J. Muszy\'{n}ska, D. Papierowska,
J. Pipin, and W. Wo\'{z}nicki,  
Int.\ J.\ Quantum Chem.\ {\bf 22}, 1153 (1982).
\bibitem{pipinwo} J. Pipin and W. Wo\'{z}nicki,
Chem.\ Phys.\ Lett.\ {\bf 95}, 392 (1983).
\bibitem{pouchan} C. Pouchan and D. M. Bishop,
Phys.\  Rev.\ A {\bf 29}, 1 (1984).
\bibitem{muller} W. M\"{u}ller, J. Flesch, and W. Meyer,
J.\ Chem.\ Phys.\ {\bf 80}, 3297 (1984).
\bibitem{knowles} P. J. Knowles and W. J. Meath,
Chem.\ Phys.\ Lett.\ {\bf 124}, 164 (1986).
\bibitem{maroulis} G. Maroulis and A. J. Thakkar,
J.\ Phys.\ B {\bf 22}, 2439 (1989).
\bibitem{Pono} D. V. Ponomarenko and A. F. Shestakov,
Chem.\ Phys.\ Lett.\ {\bf 210}, 269 (1993).
\bibitem{wang} Z-W Wang and K. T. Chung,
J.\ Phys.\ B {\bf 27}, 855 (1994).
\bibitem{merawa} M. M\'{e}rawa, M. R\'{e}rat, and C. Pouchan,
Phys.\  Rev.\ A {\bf 49}, 2493 (1994).
\bibitem{kassimi} N. E. Kassimi and A. J. Thakkar,
Phys.\  Rev.\ A {\bf 50}, 2948 (1994).
\bibitem{laughlin} C. Laughlin, J.\ Phys.\ B {\bf 28},
L701 (1995).
\bibitem{molof} R. W. Molof, H. L. Schwartz, T. M. Miller,
and B. Bederson, Phys.\  Rev.\ A {\bf 10}, 1131 (1974).
\bibitem{margoliash} D. J. Margoliash and W. J. Meath,
J.\ Chem.\ Phys.\ {\bf 68}, 1426 (1978).
\bibitem{glover} R. M. Glover and F. Weinhold,
J.\ Chem.\ Phys.\ {\bf 66}, 191 (1977).
\bibitem{bartolotti} L. J. Bartolotti, 
J.\ Chem.\ Phys.\ {\bf 73}, 3666 (1980).
\bibitem{rerat} M. R\'{e}rat, M. Caffarel, and C. Pouchan,
Phys.\  Rev.\ A {\bf 48}, 161 (1993).
\bibitem{mkchen} M.-K. Chen, J.\ Phys.\ B {\bf 28}, 4189 (1995).
\bibitem{stacey} G. M. Stacey and A. Dalgarno,
J.\ Chem.\ Phys.\ {\bf 48}, 2515 (1968).
\bibitem{manakov} N. L. Manakov and V. D. Ovsiannikov,
J.\ Phys.\ B {\bf 10}, 569 (1977).
\bibitem{bussery} B. Bussery and M. Aubert-Fr\'{e}con,
J.\ Chem.\ Phys.\ {\bf 82}, 3224 (1985).
\bibitem{konowalov} D. D. Konowalov and J. L. Fish,
Chem.\ Phys.\ {\bf 77}, 435 (1983).
\bibitem{vigne} F. Vign\'{e}-Maeder, Chem.\ Phys.\
{\bf 85}, 139 (1984).
\end{thebibliography}
\end{document}